\documentclass[twocolumn,showpacs,preprintnumbers,amsmath,amssymb]{revtex4}

\usepackage{amsmath}
\usepackage{amsfonts}
\usepackage{amssymb}
\usepackage{graphicx}
\usepackage{subfigure}
\usepackage{graphicx}
\usepackage{dcolumn}
\usepackage{bm}
\def\be{\begin{equation}}
  \def\ee{\end{equation}}
\def\bea{\begin{eqnarray}}
\def\eea{\end{eqnarray}}

\def\n{\nonumber}
\def\l{\label}
\def\p{\phi}
\def\o{\over}
\def\R{\rho}
\def\pa{\partial}
\def\om{\omega}
\def\na{\nabla}
\def\P{\Phi}

\begin{document}


\title{(Non-) geodesic motion in  chameleon   Brans Dicke model  }

\author{Kh. Saaidi}
 \email{ksaaidi@uok.ac.ir}
 \altaffiliation[Also at ]{ksaaidi@phys.ksu.edu}


  \affiliation{Department of Physics, Faculty of Science, University of Kurdistan,  Sanandaj, Iran.\\
  Department of Physics, Kansas State University, 116 Cardwell Hall, Manhattan, KS 66506, USA.}



\date{\today}

\begin{abstract}
Based on  \cite{16}, we assume there is a non-minimal coupling between the scalar field and matter in Brans-Dicke model. We analyzes the motion of different matter such as, massless scalar field, photon, massless perfect fluid (dust), massive perfect fluid and point particle matter  in this theory. We show that the motion of  massless scalar field and photon can satisfy  null geodesic motion only in high frequency limit. Also we find that the motion of the  dust and massive perfect fluid is geodesic for $L_m=-P$ and   is non-geodesic for $L_m=\R$. Finally, we study the motion of point particle and show that the motion of this kind of matter is non-geodesic.

\end{abstract}

\maketitle

\newpage

\section{Introduction}
Contemporary cosmology is encountered  with the important
challenge of understanding the existence and nature
of the dark energy  component of the Universe.  Analysis of
cosmological observations suggests that  about $\% 74$ of the Universe is dark energy (DE),  $\% 22$  is dark matter (DM) and  the remaining part is ordinary matter{\cite{1}}.
  Although the nature and origin of DE are unknown for
researchers until now, there are many proposals to explain the role
of DE to explain the accelerating expansion of the Universe. It seems that the best one is cosmological
constant$,\Lambda,$ which has the equation of state (EoS) parameter
$\omega=-1$, \cite{2, 3, 4},  and the second popular candidate of DE model are the scalar field models  with a dynamical equation of state.
The most important   dynamical DE model  which is called "quintessence" model consider the slow-roll down of a scalar field.  Although, the quintessence scalar field cannot satisfy the local tests (solar system constraints), but it  suggests an energy form with negative pressure to explain the accelerating expansion of the Universe \cite{5,6,7,8,9,10,11,12}.

Another suitable framework to investigate the behavior of DE is chameleon mechanism. In this mechanism the scalar
field has  non-minimal coupling with  matter. Chameleon mechanism provides an alternative mechanism for circumventing the constraints
from local tests of gravity. In this mechanism the scalar field acquires a mass whose magnitude depends on the local matter density.
Indeed this mechanism is a way to give an effective mass to
a light scalar field via field self interaction and interaction between field and matter {\cite{13,14}. So because of this fact  the correction of physical quantity in Newtonian regime is small, i. e., the local tests are satisfied.

Another model that has attracted much attention  is Brans-Dicke (BD) theory.  Although BD theory proved useful for  solution of many cosmological problems, but it has a problem. Indeed, the Brans-Dicke parameter, $\om$, takes small value  ($\sim 1$) when standard BD model is used
to derive the cosmic acceleration, and  on the other hand,  local constraints require that  $\om >10^4$. So, some researchers such as, Clifton \textit{et al}.,
{\cite{15}} and Das \textit{et al}., {\cite{16}}, have studied  another
framework which scalar field has non-minimal coupling with both geometry and matter. This model is called chameleon Brans-Dicke (CBD) model and it has predicated a value for $\om$ which is in a good agreement with observational data \cite{17}.

 As mentioned earlier, there are a lot of attempts to  explain  the positive accelerating expansion of the Universe. In order to do that, people have introduced various  models  for example; CBD model. Actually CBD model is studied in detailed on large scale, solar system scale and it is shown that the obtained results  are in a good agreement with observations \cite{17',18,19,20,21,22}.

However when a scalar field interacts with other components of matter
(visible matter and invisible matter \footnote{Invisible matter is not completely cold dark matter because
some candidates of cold dark matter are visible.}) through gravity or directly, this interaction may be produce a fifth
force on the matter which may violate  the weak equivalence principle (WEP) and  creates a non-geodesic motion.
This kind of interactions have attracted much attention \cite{23}. There are some particular mechanism for circumventing
the fifth force effects. Some researchers believe that the scalar field coupled differently to visible
and invisible matter of the Universe \cite{24,25,26}. Therefore based on this opinion, for suppressing the effects of fifth
force, they assume that the scalar field couples only to the invisible matter \cite{25}. Another mechanism for circumventing
the fifth force and then the violation of the WEP, is chameleon mechanism. As was mentioned earlier,
the mass of chameleon scalar field is a function of local density and  in the high density regions\footnote{Where observation and experiments are performed such as Earth.} the fifth force effects are confined to an undetectable small distances. Therefore the violation of WEP is not observed. The WEP violation had become the hot topic and studied in ditail \cite{13,27}. But there are still
another aspects of CBD model which should be studied and to the best of our knowledge, there is no any detailed
study about geodesic and non-geodesic motion for different kinds of matter in CBD model.

 This paper is organized as follows. In Sec.\;II, we consider the model, then we will obtain the equations of motion and conservation relation for density  energy. In Sec.\;III, we will study the motion of massless scalar field and photon. In Sec.\;IV we consider  the motion of perfect fluid and point particle  in this model.  At last we summarize  our work and give some discussion in Sec.\;V.

Throughout this paper, the metric signature $(-, +, +, +)$ and the convention $8\pi G =1$ are used.
\section{General Framework}
 We begin with  the chameleon Brans-Dicke action \cite{16}
\begin{equation}\label{1}
A=\int d^4x\sqrt{-g}\left[\phi R-
\frac{\omega}{\phi}\na_{a}\phi\na^{a}\phi-V(\phi)-2f(\phi){L}_{m}\right],
\end{equation}
where R is the Ricci scalar curvature, $g$ is  determinant of metric, $\omega$ is the dimensionless CBD parameter,
${L}_{m}={L}_{m}(\psi, g_{ab})$ is the Lagrangian of the matter and $\psi$ is the matter field. $\phi$ is the CBD
scalar field with a potential $V(\phi)$.
Note that the last term in the action indicates the interaction between the matter and an  arbitrary function  of scalar field, $f(\phi)$. 
 
 The gravitational field
equation can be  derived by taking account  variation of (\ref{1}) with respect to $g_{ab}$ and is given by 
\begin{equation}\label{2}
\p G_{ab} +\Big[g_{ab}\square -\na_{a}\na_{b}\Big]\p = f(\p) T_{ab} +  T^{\p}_{ab},
\end{equation}
where
\be\label{3}
T_{ab}=  {2\o \sqrt{-g}}{\delta (\sqrt{-g}L_m)\o\delta g^{ab}},
\ee
is the the matter energy-momentum tensor and
\be\label{4}
T^{\p}_{ab} = {\om \o \p}\Big [ \na_{a}\p\na_{b}\p - {1\o2}g_{ab} \na_{\alpha}\p\na^{\alpha}\p\Big] - {1\o 2} g_{ab} V(\p),
\ee
is the scalar field energy-momentum tensor.  Also by taking the variation of (\ref{1}) with respect to $\p$ we have the equation of motion for scalar field
\be\label{5}
(3 + 2 \om)\square \p = f(\p) T -  \p f'(\p) L_m + \p V'(\p) -2 V(\p).
\ee
 Here $T$ is the trace of matter energy-momentum tensor and prime denotes derivative with respect to $\p$. It is seen that to solve  (\ref{5}) we need an explicit form of matter Lagrangian, $L_m$.

  The Bianchi identities, together with the identity
$(\square\nabla_a - \nabla_a\square ) V_c = R_{ab}\nabla^bV_c$, imply the non-(covariant)
conservation law
\be\label{6}
\nabla_aT^{ab} = -\big[g^{ab}L_m + T^{ab}\big]{\nabla_a \ln(f)},
\ee
and, as expected, in the limit $f(\p)$ = constant, one recovers the
conservation law $\nabla_aT^{ab} = 0$.
Since the energy-momentum tensor is not covariantly
conserved, one may  concludes that the motion of matter distribution
characterized by a Lagrangian density $L_m$ is nongeodesic.
 This  fact that the  energy-momentum tensor is not    divergence-free, can be interpreted as a violation
of the so-called metric postulates \cite{28}.

\section{ Matter-scalar coupling and geodesics of massless matter field}

As was mentioned earlier in the Introduction, the explicit coupling between matter and  scalar field which is  described by the
action (\ref{1}) can potentially lead to non-geodesic motion. In this section we consider the affect of this kind of interaction on  massless particles geodesics.

\subsection{Massless scalar field}
Let us consider  a massless scalar field $\psi$ which is described by the Lagrangian density
 \bea\label{7}
 L_m= -{1\o2} \nabla_a\psi\nabla^a\psi.
 \eea
 Using
 $$T_{ab} = {\pa L \o \pa (\na^a\psi)}\na_b\psi - g_{ab}L,$$
 one can obtain the stress-energy tensor of the scalar field
 \bea\l{8}
 T_{ab}= -\nabla_a\psi\nabla_b\psi+ {1\o2}g_{ab}\nabla_c\psi\nabla^c\psi, \l{9}
 \eea
 Substituting (\ref{7}) and (\ref{8}) in (\ref{6}) we have
 \be\l{9}
\square\psi = - \nabla_a\psi\nabla^a\ln(f),
 \ee
 One can see that for $L_m$ and $L_m+\na_c\chi$ ($\chi$ is a scalar function),
Eq.\;(\ref{9})) is not changed just for $\partial\chi_{,c}/\partial\na^a\psi = \na_a\chi_{,c}=0$,
but equation of motions for other components of the system, Eqs.\;(\ref{5}) and (\ref{6}), are changed in this case. This
means that when massless scalar matter couple with other components of the system, there is no any degeneracy
of Lagrangian densities.

 Note that this kind of matter is known as "{\it scalar photon}".
So, although our study is completely classic, but we assume this scalar photon has a wave like behavior,
then the matter scalar field, $\psi$, can be as a wave function. Therefore we assume the wave function is a high
frequency wave as
  \be\l{10}
 \psi(x) = \psi_0 e^{i\Phi(x)}.
 \ee
 Here the phase of wave is a rapidly varying function of $x$  and  $\psi_0$ is nearly constant. Therefore Eq.(\ref{8}) becomes
 \be\l{11}
 i\square \Phi(x) - \nabla^a\Phi\nabla_a \Phi= -i\nabla_a\Phi\nabla^a\ln(f).
 \ee
   Since this equation has tow  real and imaginary parts, we have
 \bea\l{12}
  \nabla^a\Phi\nabla_a\Phi &=& 0,\\
   \square\Phi(x) &=& -\nabla_a\Phi\nabla^a\ln(f),\l{13}.
   \eea
In comparison with similar analysis in standard model of cosmology, Eq.\;(\ref{13}) shows that the scalar particle is
not transverse unless $\na_a\P$ will be orthogonal to $\na^a\ln(f)$ or $\na^a\ln(f)=0$. Also by taking covariant derivative of
Eq.\;(\ref{12}) and using $\na_a\na_b\P = \na_b\na_a\P$, one can obtain
\be\l{14}
\na_v v_a = 0,
\ee
 where $v_a = \na_a\P$. Note that $v_a = \na_a\P$ is the gradient of
the wave phase and in the geometric optic approximation, this quantity is the tangent of the worldline of the particle
(massless scalar particle). Therefore in comparison to real photon, Eq.\;(\ref{14}) means that the motion of massless
scalar particle in the high frequency limit is take place on the null geodesic.

\subsection{Maxwell field}
Let us consider the Maxwell field with Lagrangian density and energy- momentum tensor
\bea\l{15}
L_m&=& -{1\o 16\pi} F^2,\\
T_{ab} &=& -{1\o 4\pi} \bigg[ F_{ac}F_b^c - {1\o 4}g_{ab} F^2\bigg],\l{16}\eea
where $F_{ab}$ is the electromagnetism field tensor on the curve space-time
\be\l{17}
F_{ab} = \na_aA_b-\na_bA_a = \pa_aA_b-\pa_bA_a,
\ee
 and $A_a$ is the vector potential. Substituting (\ref{15}) and (\ref{16}) into (\ref{6}) gives
 \be\l{18}
 \na^b\bigg[ F_{ac}F_b^c - {1\o 4}g_{ab} F^2 \bigg] =- F_{ac}F_b^c\na^b(\ln f).
 \ee
 Now we consider the high frequency limit. For this goal we introduce the vector potential as
 \be\l{19}
 A_a(x) = C_a e^{i\Phi(x)}
 \ee
  where $C_a$ is a slowly varying vector amplitude (nearly constant) and $\Phi(x)$, the phase of wave function, is a rapidly varying function.
  So by neglecting the derivatives of vector amplitudes, $C_a$, we have
  \bea\l{20}
 2 {\cal A}_{ab} \na^b\Phi  - \na_a\Phi\bigg[ \big(\na\Phi\big)^2C^2 - \big( \na_c\Phi C^c\big)^2 \bigg]=0,
 \eea
 \bea\l{21}
 2\na^b\Big[{\cal A}_{ab}f(\p)\Big] -\na_a\bigg[ \big(\na\Phi\big)^2C^2 - \big( \na_c\Phi C^c\big)^2 \bigg]=0,
 \eea
 where
 \bea\l{22}
 {\cal A}_{ab}&=& C^2\na_a\Phi\na_b\Phi + \big(\na\Phi\big)^2C_a C_b \n \\ &&- \big(C^c\na_c\Phi\big)\Big[C_b\na_a\Phi+ C_a \na_b\Phi\Big],
 \eea
 and $C^2 = C^cC_c$ , $\big(\na\Phi \big)^2 = \na_c\Phi\na^c\Phi$.
 Using Eqs.\;(\ref{20}) and (\ref{22}), one can find
 \be\l{23}
 C^2\big( \na\Phi\big)^2 = \big(C^c\na_c\Phi\big)^2,
 \ee
 and by substituting (\ref{23}) into (\ref{20}) and (\ref{21}) we have
 \bea\l{24}
 {\cal A}_{ab} \na^b\Phi &=&0,\\
 \na^b{\cal A}_{ab} &=&- {\cal A}_{ab}\na^b\ln(f).\l{25}
 \eea
 By setting $f(\p) =1$ one can arrive at the standard Maxwell equations in curved space. Geometric optics is valid whenever the wavelength
 is very short with respect to the radius of curvature of space-time, namely $\bar{\lambda} \ll {\cal L}$, here ${\cal L}$ is the radius of curvature of space-time, and $\bar{{\lambda}}$ is the reduced wave length of photon.   One can write Eq.\;(\ref{25}) as
 \be\l{26}
 |\na^b{\cal A}_{ab}|\simeq|{{\cal A}_{ab} \o \bar{\lambda}}|\gg |{\cal A}_{ab}{\ln(f)\o {\cal L}}|
 \ee
So, in this case  we have
 \bea\l{27}
  \na^b{\cal A}_{ab} \simeq 0.
 \eea
 This shows that the corrections to standard optics which is coming from interaction between scalar field and matter character, $f(\p)$,  complectly removed from the equation of motion and then photons follow the null geodesics and are transverse. On the other hand for the case that  $\bar{\lambda} /{\cal L} \nless 1$, one can not disappear the non-minimal coupling affect to the Maxwell equations and then the null  geodesic equation of photon is modified.

\section{Matter-scalar field coupling and geodesics of perfect fluid matter}
In this section we consider a kind of matter, so called perfect fluid, which can be massive or massless.  The stress-energy tensor of perfect fluid is represented by
\be\l{28}
T_{ab}= (\R + P)U_aU_b + g_{ab}P,
\ee
    where $\R$ is the energy density and $P$ is the pressure of the matter respectively, and the four velocity, $U_a$ satisfies the constraints $U_aU^a = -1$ and $U^a\na_bU_a=0.$ In \cite{29,30,31,32,34} have been shown that, for perfect fluid that does not couple explicitly
to the other components of the system, there are different Lagrangian densities which are perfectly equivalent.
In fact, they have shown that, by using Eq.\;(\ref{3}), the two Lagrangian densities $L_{m_1} = -P $ and $L_{m_2} = \R$
give the same stress-energy tensor as (\ref{28}), and also for these two different Lagrangian densities the equation of
motions for all components of the system is similar. Also, since the perfect fluid laws are obtained via a kinetic theory
by using microscopic models of the fluid particles and  their interaction, namely, the perfect fluid is an averaged
and not an exact description for matter, it is more common to work directly with the energy-momentum tensor
instead of Lagrangian density in the non-interacting model of perfect fluid. But in our model the Lagrangian
density, $L_m$, is explicitly appeared in equation of motion of scalar field, (\ref{5}), and conservation relation, (\ref{6}). So we
encounter with a new situation that we have to study it accurately.

 We can work with stress-energy tensor of perfect fluid only for the case which there is no any interaction between
perfect fluid and other components of the system. This means that if there is a direct interaction between
perfect fluid and other components of system, such as geometry and scalar field, the above Lagrangian densities
give rise to distinct theories with different predictions. To show this fact, let us consider a general case.

We assume there is a minimal coupling between  perfect fluid and scalar filed, i.e., $f(\p) = 1$.
In this case the Lagrangian of perfect fluid is not appeared explicitly in the equation of motions of other components
of the system and the energy-momentum tensor of matter is conserved, namely
\bea\l{29}
\p G_{ab} + \big[g_{ab}\square - \na_a\na_b \big ]\p &=& T_{ab} + T^{\p}_{ab},\\
(3 + 2\om)\Box\p - \p V'(\p) +2 V(\p) &=& T, \l{30}\\
\na_aT^{ab} &=& 0, \l{31}
\eea
where matter energy-momentum tensor, $T_{ab}$, and the scalar field energy-momentum tensor, $T^{\p}_{ab}$, are given by
Eqs.\;(\ref{3}) and (\ref{4}). Also we suppose there are two different Lagrangian density $L_{m_1}$ and $L_{m_2}$ which by using (\ref{3}) give
an energy-momentum tensor like (\ref{28}), and $L_{m_1}$ and $L_{m_2}$ are related with together by
\be\l{32}
L_{m_2} = L_{m_1} + {1\over \sqrt{-g}} \na_a\chi
\ee
where $\chi$ is a scalar function. Since the perfect fluid Lagrangian is not appear in equation of motions, (\ref{29}) and
(\ref{30}) and conservation relation of energy, (\ref{31}), then these two Lagrangian has not any effect on equation of motion
of other components of the system and conservation relation of energy. Also the additional term, $\na_a\chi/\sqrt{-g}$,
in Eq.\;(\ref{1}) give a surface integral term, then this term has no any effect on equation of motions of perfect fluid.
Therefore these two different Lagrangian are equivalent for perfect fluid in a non-interacting model.

On the other hand we assume, there is an interaction between perfect fluid and other components of system,
i.e., for an arbitrary $f(\p)$. It is obviously seen that the equation of motion of perfect fluid for two different Lagrangian
densities, (\ref{32}), are similar, moreover, by using Eq.\;(\ref{3}) one can obtain a matter energy-momentum tensor
as (\ref{28}). But the equation of motion of scalar field and also the conservation relation of energy for matter
become equations (\ref{6})and (\ref{7}) respectively, which are different with Eqs.\;(\ref{29}) and (\ref{30}) and they  are different for $L_{m_1}$
and $L_{m_2}$. This fact shows that, clearly, the two Lagrangian density $L_{m_1}$ and $L_{m_2}$ cannot be equivalent in
an interacting system of perfect fluid with other components of the system.

\subsection{Null geodesic of dust}
The  equation of null geodesics for a model with out any coupling  is  derived  from the conservation equation of a null dust fluid in   \cite{35}.
Therefore, in this case which there is a coupling between scalar field and matter we do the same way of derivation. If this interaction  were to induce any corrections to the null geodesic equation, these has to  show up in this derivation. Since dust is a perfect fluid without pressure, namely $P=0$ then the setters-energy is $T_{ab}=\R U_aU_b$, so by using the modified conservation equation, (\ref{7}), we have
\be\l{33}
\na_uU_a = \eta U_a  ,
\ee
where $\na_u = U^b\na_b$ and
\be\l{34}
\eta = -\na_u\ln(f\R) -\na^bU_b,
\ee
Eq.\;(\ref{33}) is a geodesic equation which is non-affinely parameterized. In fact this equation  shows that the four velocity is transported along the path parallel to itself and this is the definition of a geodesic curve. So this means that the existence of coupling between the matter and scalar field does not change the equation of null geodesic for $L_m= -P$. But for $L_m= \R$ we obtain
\be\l{35}
\na_uU_a = \eta U_a - \na_a\ln(f),
\ee
where clearly shows that parallel transport is no longer
conserved and then the motion of dust particle is non-geodesic in this case.

\subsection{Massive perfect fluid matter}
In this section we  turn our attention to massive matter fields and, for simplicity, consider a perfect fluid composed of non-relativistic or relativistic particles with stress-energy tensor (\ref{28}). This stress-energy tensor  is obtained from (\ref{3}) with $L_m = -P$. By defining a projection operator as $h_{ab} = g_{ab} + U_aU_b$, one can project equation (\ref{6}) onto the direction normal to the four velocity as
\be\l{36}
\na_bT^{ab} = -\big[ g_{ab}L_m + T_{ab}\big] \na^b(\ln f).
\ee
Using (\ref{28}), one can obtain the non-geodesic motion for the fluid element as
\be\l{37}
\na_uU^c = f^c,
\ee
where the extra force, $f^c$ is given by
\be\l{38}
f^c = -{1\o (P+\R)} \bigg[ (L_m+P)\na_b(\ln f) + \na_bP\bigg]h^{ac}.
\ee
So by using $L_m = -P$ we have
\be\l{39}
f^c = -{1\o (P+\R)}  \na^cP.
\ee
This states the extra force is related  to the coupling between matter and scalar field and it is  proportional to the pressure gradient. This term  is the usual term that appears in standard  GR and encapsulates the force exerted on a fluid element due
to the fluid pressure. This means that  energy is indeed not conserved for
this fluid does not affect  the geodesic motion in GR.

Also by substituting (\ref{28}) in to (\ref{37}) one can obtain
\be\l{40}
\na_aT^{ab} = -(P+\R)\big[\na_u(\ln f)\big]U^b,
\ee
this equation states
 the flow of energy only take places  along the direction of $U$ i.e., aligned with the fluid worldlines.  This means that  the spatial components of the force in the rest frame of the fluid is zero and only the time component of the force is nonzero.
 This kind of force cannot  have any  effect on the motion because, based on the  normalization $U^aU_a = -1$, the four acceleration $a_c$ is perpendicular  to the four velocity $U^c$. This states that the components of four-force perpendicular to the four velocity is zero. This is the fact which we discover here.
Note that according to (\ref{38}) in the case of dust with $P=0$, the extra force $f^c$ is zero and this is agree with (\ref{33}).

On the other hand by inserting the Lagrangian density $L_m = \R$ in Eqs.\;(\ref{36}) and (\ref{37}) we have
\bea\l{41}
\na_aT^{ab} &=& -(P+\R)\big[\na^b(\ln f)\big],\\
f^c &=& -{1\o (P+\R)}  \na^cP - \na^c\ln(f).\l{42}
\eea
Equation (\ref{41}) states the flow of energy is not along the
direction of $U$. This means that all components ( spatial
and time components ) of the fifth force in the rest frame
of the fluid are nonzero and then this kind of force can
have any effect on the motion of matter. Also Eq.\;(\ref{42})
shows that the fifth force does not proportional to the
pressure gradient and it depends to the coupling between
matter and scalar field. This means that in this case the
motion is not geodesic.

\subsection{ Massive matter}
In this Subsection we want to study the motion of ordinary massive matter (not perfect fluid). For this goal, we begin
 with energy-momentum four-vector $p^a_0(t)$. The density of $p^a_0(t)$ is defined by \cite{36}
\be\l{43}
T^{a0}(x,t) = p^a_0(t)\delta(\mathbf{x}-\mathbf{x_0}),
\ee
where $\mathbf{x}$ is the general coordinate, $\mathbf{x_0}$ is the coordinate of
center of particle ( the index "$0$" indicates the center of particle). And the current of this four vector is defined
by
\be\l{44}
T^{ai}(x,t) = p^a_0(t){dx^i_0(t)\over dt}\delta(\mathbf{x}-\mathbf{x_0}),
\ee
one can united these two definition into one as
\be\l{45}
T^{ab}(x,t) = p^a_0(t){dx^b_0(t)\over dt}\delta(\mathbf{x}-\mathbf{x_0}),
\ee
where $x^0_0(t) := t$. By rewriting the energy-momentum tensor of particle, (\ref{45}), in the  co-moving coordinate and using
$p^a_0(\tau) = m_0u^a$, we have
\be\l{46}
T^{ab}(x) = m_0u^au^b,
\ee
where $\tau $  is the proper time, $u^a$ is four velocity vector and $u^au_a = -1$.

Moreover according to \cite{26}, we introduce a matter Lagrangian
for a point particle with mass $m_0$ by
\be\l{47}
L_m= m_0\delta(\mathbf{x}-\mathbf{x_0})\sqrt{-g_{ab}\dot{x}^a_0\dot{x}^b_0},
\ee
where $\dot{x}^a_0 = dx^a_0/d\tau$. The Lagrangian (\ref{47}) give the particle equation of motion as
\be\l{48}
\ddot{x}^a_0 + \Gamma^a_{bc}\dot{x}^c_0\dot{x}^b_0 = 0,
\ee
This  equation is  a geodesics equation of motion for point particle. Since $g_{ab}\dot{x}^a_0\dot{x}^b_0 = g_{ab}u^au^b= -1$,  one can rewrite (\ref{47}) in co-moving coordinate  as
\be\l{49}
L_m = m_0.
\ee
Substituting Eqs.\;(\ref{46}) and (\ref{49}) into Eq.\;(\ref{6}) we get
\be\l{50}
\na_uu_b = - \tilde{\na}_b\ln(f) -(\na_au^a)u_b.\ee
Eq.\;(\ref50}), is  geodesic equation of motion for a particle in CBD model which is modified with respect to conventional
geodesic equation of GR. There are two terms in the right side of Eq.\;(\ref{50}) which are the fifth force contribution
from non-minimal coupling between matter and scalar field. Note that $\tilde{\na}_b := (g_{ab} + u_au_b)\na^a$ is a particular
derivative in the 3-dimensional space perpendicular to $u_a$. Therefore $\tilde{\na}_b\ln(f)$ is perpendicular to $u_a$,
so it doesn't any effect on the magnitude of velocity, namely this term does not any effects on the energy of
particle. Also, since $(\na_au^a)u_b$ is aligned on the particle worldline, it does not any effects on parallel transport of
four-velocity.

\section{Conclusion}
We have studied Brans-Dicke model which include a non-minimal coupling between scalar field and matter,
so-called chameleon Brans-Dicke model. We have considered the possible deviation of free fall trajectories from
geodesics. We have studied the motion of massless scalar particles, photon, massless perfect fluid(dust), massive
perfect fluid and finally ordinary massive particles.

By assuming a wave like behavior for massless scalar
field ( scalar photon), we have shown that the motion  of scalar photon is take place on null geodesics only for
high frequency limit, Moreover we have found that the electromagnetic particle (photon) is transverse and the
motion of it is null geodesics only for the case which  the reduced wave length of the photon be very small with
respect to radius of curvature of the space-time and foe the case $\bar{\lambda} /{\cal L} \ll 1$ the photon does not transverse and the
motion of it is not  null geodesic.

Furthermore, we have discussed the (non)- geodesics motion of perfect fluid. We have found that, although
there is a degeneracy of Lagrangian densities in the context of standard GR, but our analysis have shown that in
the CBD model this degeneracy does not excite and also dust and massive perfect fluid have geodesics motion foe
$L_m = -P$ and the motion of them is non-geodesics for $L_m =\R$.

Finally, we introduced a Lagrangian and energy-momentum tensor for a point like particle, and we have
shown that the motion of a massive particle is not geodesic in CBD theory.
\section{Acknowledgement}
 The work of  Kh. Saaidi has been supported financially  by the University of Kurdistan, Sanandaj, Iran, and  he would like thank to the University of Kurdistan for supporting him in his sabbatical period.

\end{document}